# Prior Knowledge Helps Improve Beam Angle Optimization Efficiency in Radiotherapy Planning

Yong-Jie Li, *Senior Member, IEEE*

*Abstract*—One of the grand challenges in intensity-modulated radiotherapy (IMRT) planning is to optimize the beam angles in a reasonable computation time. To demonstrate the value of prior knowledge on improving the efficiency of beam angle optimization (BAO) in IMRT, this study proposes a modified genetic algorithm (GA) by incorporating prior knowledge into the evolutionary procedure. In the so-called nowGABAO (kNOWledge guided GA for BAO) approach, two types of prior knowledge are incorporated into the general flowchart of GA: (1) beam angle constraints, which define the angle scopes through which no beam is allowed to pass, and (2) plan templates, which define the most possible beam configurations suitable to the studied case and are used to guide the evolutionary progress of GA. Optimization tasks with different prior knowledge were tested, and the results on two complicated clinical cases at the head-and-neck and lung regions show that suitable defining of angle constraints and plan templates may obviously speedup the optimization. Furthermore, a moderate number of quite bad templates produce negligible influence on the degradation of optimization efficiency. In conclusion, prior knowledge helps improve the beam angle optimization efficiency, and can be suitably incorporated by a GA. It is resistant to the inclusion of bad templates.

*Index Terms*—Intensity-modulated radiotherapy (IMRT), beam angle optimization (BAO), genetic algorithm (GA), prior knowledge.

## I. Introduction

INTENSITY-modulated radiotherapy (IMRT) is now becoming a powerful technique for the treatment of malignant tumors. To achieve acceptable IMRT planning, many optimization algorithms have been proposed to optimize the beam angles [2-24], [38-56] or the deliverable segments [57-65]. However, in many clinical practices, the conventional IMRT planning starts with the manual selection of several suitable beam angles, followed by an optimization of beam intensity maps using inverse optimization methods [1]. The clinical requirement to automatically select beam angles is particularly highlighted in IMRT, by which highly three-dimensionally conformal dose distribution could be achieved with intensity-modulated beams irradiating from different directions. During the past three decades, the importance of selecting suitable beam angles for IMRT treatment has been widely demonstrated, and extensive efforts have been made to facilitate the technique of automatic beam angle selection in IMRT practice [2]–[24]. Though there are fruitful results achieved, the improvements are still not quite satisfying, especially on the clinically infeasible computation time resulted from the huge hyperspace of solutions in beam angle selection. To date, defining suitable beam angles is still one of the grand challenges in IMRT planning.

Besides further refining the existing algorithms and developing new methods, the idea of utilizing certain prior knowledge to help the computer improve beam angle optimization efficiency has emerged in several works in order to tackle this time-consuming task. Rowbottom *et al.* used an artificial neural network (ANN) algorithm to predict beam orientations in conformal radiotherapy [8]. Xing *et al.* developed a medical knowledge based system to facilitate the selection of beam orientations based on previously calculated skeletal plans for a selection of patients [9]. Subsequently, Xing and his colleagues introduced two measures, namely, beam's-eye-view dosimetrics (BEVD) [14] and EUD-based function [25], as prior knowledge to evaluate automatically the quality of possible beam orientations. Recently, Vaitheeswaran *et al.* defined a measurement of "beam intensity profile perturbation score (BIPPS)" to rank beam angles [34]. These angular ranking metrics could be used either as prior information to speedup the computer to search for the optimal beam configuration by guiding the optimization progress, or as a guidance to facilitate the manual beam selection.

Besides the prior information utilized in the above-mentioned works on facilitating the beam angle optimization, another kind of knowledge associated with human experience should also gain sufficient attentions. In fact, plentiful prior knowledge has been accumulated by oncologists and physicists over the time during their clinical practice. For examples, some beam angles cannot be used in clinical practice because of the physical limitation or potentially severe danger for treatment. Normally, there are three main situations for beam angle restriction: (1) beam orientations that would result in a collision between the treatment gantry and patient couch; (2) beam orientations passing through the organs-at-risk (OARs) that have a extra low radiation tolerance, and (3) beam orientations directly passing the metal frames of the patient couch or immobilizing couch that

Y. Li is with the School of Life Science and Technology, University of Electronic Science and Technology of China, Chengdu 610054, China (e-mail: liyj@uestc.edu.cn). .

can normally attenuate the radiation. Furthermore, with the accumulation of clinical experience, some plan configurations have become informal standards for the treatment of specific tumors in some institutions, and some standard solutions have been found for some typical tumors [24], [29]. It is natural for planners to expect optimization results to be more satisfying within acceptable computation time by fully utilizing their prior knowledge mentioned above. Recently, Yuan *et al.* investigated the feasibility of establishing a small set of standardized beam bouquets for lung IMRT planning [54], and their experiments show that appropriately standardized beam configuration bouquets can indeed help improve plan efficiency and facilitate automated planning.

On the basis of the general flowchart of our previously developed genetic algorithm (GA) based beam angle optimization method [20], this paper proposes a new efficient technique called nowGABAO (kNOWledge guided GA for Beam Angle Optimization) to improve the efficiency of IMRT beam angle optimization by incorporating prior knowledge into the optimization procedure. In the proposed nowGABAO, prior knowledge is used to guide GA to select beam angles in a more efficient way, and the beam intensity maps of these angles are optimized using a conjugate gradient (CG) method [20], [35]. Besides clarifying the details on the usage of prior knowledge, this paper gives special focus on a detailed analysis of the influence of prior knowledge, especially the plan templates, on the optimization efficiency on two typical clinical cases.

## II. MATERIALS AND METHODS

### A. About the Beam-Angle-Optimization Problem

In IMRT, each external beam irradiating from the gantry head of the linac is divided into a group of beamlets, the intensities of which are to be optimized in order to deliver prescribed doses to the tumor in the body while protecting the critical organs and normal tissues as much as possible. Mathematically, beam angle optimization is a combinational optimization problem, in which a specified number of beams are to be selected among a beam-angle candidate pool. This optimization problem is further complicated by the requirement of optimizing the corresponding beamlet intensity maps of these intensity-modulated beams at the given angles. Once the optimized beamlet intensity maps are found, the corresponding dose distribution in the body is calculated and the quality of the trial plan configuration is evaluated on the basis of the objective function or fitness function and the user-defined dose constraints. In short, beamlet intensity maps are coupled with IMRT beam configurations, requiring beamlet-intensity-map optimization for every sampled beam configuration. A more detailed description of the problem of beam angle optimization in IMRT could be referred to our previous work [32].

### B. Definition of Prior Knowledge

Two types of prior knowledge about the individual treatment are employed in our method: (1) beam angle constraints (ACs),

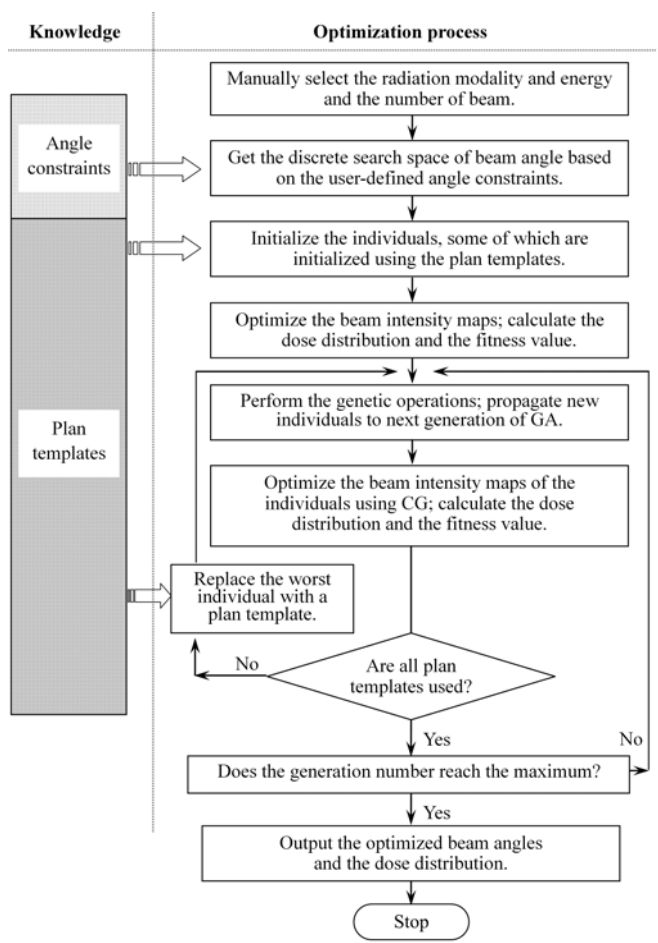

Fig. 1. The general scheme of incorporating prior knowledge into the optimization process of standard GA.

which define the angle scopes through which no beam is allowed to pass, and (2) plan templates, which are prior beam configurations potentially suitable to the studied treatment site. ACs are used to reduce the search space by excluding the defined angle scopes from the whole angle space. Plan templates are used to (1) initialize some of the individuals (i.e., treatment plans) of the first generation of GA (the remaining individuals are initialized randomly), and (2) replace the worst individual in each new generation.

### C. Knowledge-Guided GA for Beam Angle Optimization

Figure 1 demonstrates the general scheme of incorporating prior knowledge into the optimization flow of the standard GA. Our previous work has reported an automatic beam angle selection (ABAS) algorithm based on the combination of GA and CG methods [20]. The technique of nowGABAO proposed here is a substantial extension and improvement to the ideas in ABAS. The implementation of ABAS is summarized below, accompanied by a special focus on the details of incorporating prior knowledge into the optimization process of GA.

In nowGABAO, the search space covering the total 360° coplanar gantry angle is first reduced by the user-defined orientation constraints, and the remaining is then divided into equally spaced orientations with a given angle step, such as 5°. These discrete angle candidates are used to encode the new plans with trial angles. Our GA adopts a one-dimensional integer-coding scheme, in which the combination of beam angles is represented by a chromosome (i.e., an individual or a plan), and each gene in the chromosome represents a trial beam angle [20]. For example, a chromosome with five genes (i.e. a plan with five beams) could be {0°, 40°, 100°, 190°, 300°}. Note that no two genes in an individual are allowed to have the same values, which means that beams with the same angles are not allowed in a treatment plan.

In nowGABAO, two strategies are designed to initialize the individuals in the first generation of GA: (1) some ones are initialized with the defined plan templates; (2) the remaining individuals are randomly initialized. Knowledge-guided initialization of partial individuals is one of the key features of nowGABAO. It should be noted that no more than half of the total individuals in the first generation of GA are allowed to be initialized using the plan templates, in order to avoid that the prior knowledge dominates the GA operations at the very early stage of the optimization. If there are more plan templates remaining in the template list, they would be used to guide the genetic operations.

The quality of each individual is evaluated by a fitness value, and the purpose of optimization is to find the individual with the maximum fitness by minimizing the dose difference between the prescribed and calculated dose distributions. Let $B = (b_1, b_2, \cdots, b_N)$ be the beam set of an *N*-beam plan, the fitness function $Fitness(B)$ is given by [20], [32]

$$Fitness(B) = F_{\max} - F_{obj}(B) \qquad (1)$$

$$F_{obj}(B) = \sum_{i=1}^{NO} \sum_{j=1}^{NP_i} \delta_{ij} \cdot (d_{ij}(B) - p_{ij})^2 \qquad (2)$$

where $F_{obj}(B)$ is the objective function value of the plan $B$, and $F_{\max}$ is a rough estimation of the maximum value of the objective function, which is used to keep all the fitness values positive and can be simply set it to be a little greater than, e.g., 1.5 times of the maximum objective function value among the individuals in the first generation [20], [32]. $NO$ is the number of the organs involved in the treatment site. $NP_i$ denotes the number of sampling points in the *i*th organ. $\delta_{ij} = 1$ when point dose in the organ breaks the specified constraints, else $\delta_{ij} = 0$. $d_{ij}(B)$ and $p_{ij}$ represent the calculated and prescribed doses of the *j*th point in the *i*th organ, respectively.

According to the fitness values, the individuals were processed with four types of genetic operations, i.e. (1) selection, (2) crossover, and (3) mutation, and (4) immunity. Figure 2 shows the flowchart of the genetic operations. The implementation of the four genetic operations is briefly described below, and the implementation details could be referred to Ref. [20]. Parent individuals with higher fitness are selected into the next generation with a higher probability. To any two randomly selected parent individuals (i.e. plans), a crossover operation is applied according to a specified crossover probability, normally 0.5–0.95. Then a mutation operation to the two children angle sets is done according to a mutation probability, normally 0.001–0.02. Finally, a pseudo immunity operation is applied to the two children angle sets in order to enhance the genetic process.

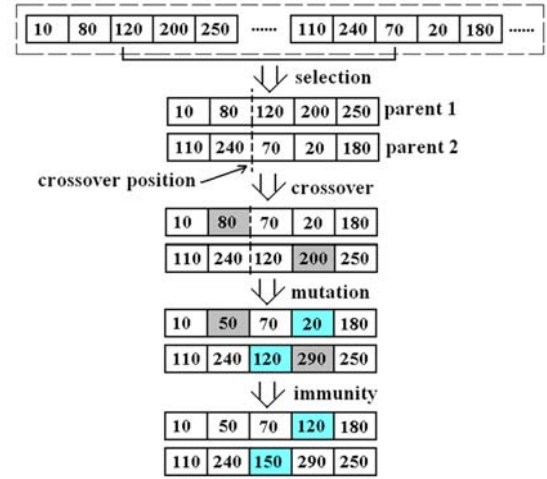

Fig. 2. The genetic operations for beam angle optimization. Parent individuals with higher fitness are selected into the next generation with a higher probability. To any two randomly selected parent individuals (angle sets), a crossover operation is applied according to a specified crossover probability. Then a mutation operation to the two children angle sets is done according to a mutation probability (e.g. the randomly selected "80" and "200" are mutated to "50" and "290", respectively). Finally, a pseudo immunity operation is applied to the two children angle sets in order to enhance the genetic process, mainly by reducing the cases that two angles in one plan are too close to each other, e.g. "10" and "20" in the first of the two chromosomes after mutation are two close for a five-beam plan, and hence, "20" is randomly changed to "110".

After the four genetic operations in each generation, the current worst individual, i.e. the one with the lowest fitness value, is replaced with a plan template if there is prior plan template remaining. Replacing the worst individual with a plan template is another key feature of the proposed method. Any prior plan template can be used only once. The optimization progress will be terminated until no plan template remains and there is no better individual found in a specified number of successive generations.

If some beam angles in a plan template come into conflict with the defined orientation constraints, these angles would be adjusted to fit the constraints. For example, given that 30°~60° is defined as an orientation constraint, which means no beam is allowed to pass through the range of 30°~60°. If beam angles in a three-beam plan template are 40°, 120° and 280°, the angle of 40° will be randomly adjusted to 30° or 60°. The basic rule of the adjustment is that the new plan template is as close as possible to the original one, under the condition of matching the defined orientation constraints.

We define that the optimization would not be terminated until no plan template remains. To avoid that too many templates would prolong the optimization unexpectedly, we tentatively select the first 50 templates as the prior knowledge if there are more than 50 plan templates.

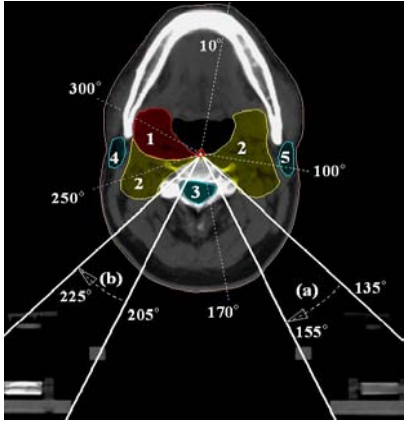

Fig. 3. The clinical head-and-neck (HN) case. There are total six volumes involved: 1 – GTV, 2 – nodes, 3 – spinal cord, 4 – right parotid, 5 – left parotid, and brainstem (not visible on this specific plane). For a five-coplanar-beam plan, the previously optimized beam angles are 10°, 100°, 170°, 250° and 300° (the light dotted straight lines). In order to avoid obstructions, two angle constraints were defined (the white solid straight lines): (a) 135°~155° and (b) 205°~225°.

## III. RESULTS

Two clinical cases at the head-and-neck and lung regions were used to evaluate the performance of nowGABAO. Tumors at such two sites are typical situations that need IMRT treatment because these tumors are normally quite close to the surrounding critical organs [1]. In order to fully test the influence of prior knowledge on optimization performance, four kinds of optimization tasks were studied for the two cases: (1) optimization without any prior knowledge, (2) optimization with only angle constraints (ACs), (3) optimization with both ACs and *good* plan templates, and (4) optimization with both ACs and *bad* plan templates. Considering that the importance of AC on improving optimization efficiency is easily understood because this type of knowledge results in an obvious reducing of the search space, here we focused more on analyzing the influence of the quality and quantity of plan templates.

The preparation of good and bad plan templates is as follows. For each tumor case, we first collected 100 approved manual IMRT plans with the same beam number as the studied case, and the top 30 IMRT plans (with the lowest objective function values that were previously computed by a CG algorithm [20]) out of the 100 templates were selected as the good plan templates for the studied case. Note that these manual IMRT plans were collected from two tumor hospitals in the city of the author and were manually designed in a trial-and-error manner using the Varian's Novalis Tx System and Elekta's Synergy IGRT System by experienced oncologists and physicists.

In addition, we produced some very *bad* plan templates by randomly defining all the beam angles of each plan template to locate within a small range of 60°. For example, a bad four-beam template could be {0°, 10°, 40°, 60°}, or {100°, 130°, 140°, 160°}, which are generally too poor to be optimal or clinically acceptable solutions.

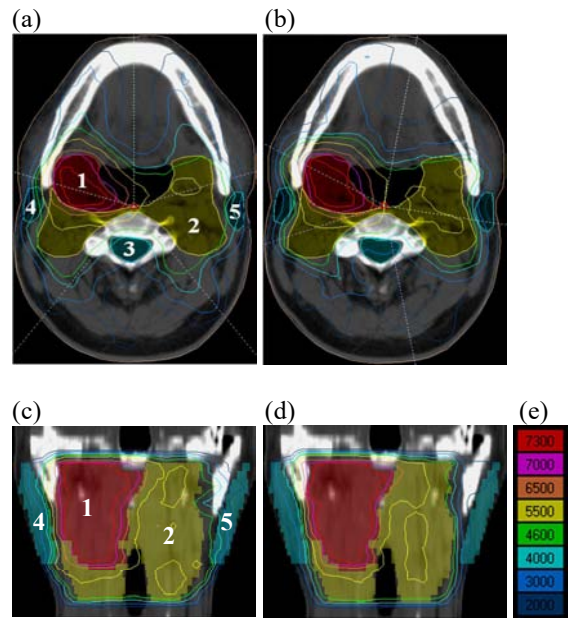

Fig. 4. Head-and-neck isodose contours on the transversal planes ((a) and (b)) and coronal planes ((c) and (d)) for the best five-beam manual plan among the plan template pool ((a) and (c)) and the nowGABAO-optimized plan ((b) and (d)). The beam angles of the manual plan are 0°, 70°, 150°, 210° and 290°. The optimized beam angles are 10°, 100°, 170°, 250° and 300°. The legend of the isodose level (color vs. dose (cGy) ) is shown in (e).

The beam angle search space was sampled with a step of 5°, which results in 72 discrete angles among the whole 360° coplanar gantry range. The population size of GA was empirically set to be 20 for the two studied cases [12], [17], [20]. All other parameter settings for GA in this study, such as crossover and mutation probability, etc., were same as that in our previous work [20]. Both the IMRT performance (in terms of dose distribution and Dose-volume histogram (DVH)) and optimization efficiency (in terms of generation number of GA) were compared between the manual and nowGABAO plans. For a fair comparison, each optimization task was run five times and the needed generation number was averaged over the five runs. Note that for each run, the good templates were randomly selected from the total 30 good templates mentioned above.

Before running nowGABAO, we first employed an exhaustive approach to find out the optimal beam angles for the two studied cases under the condition that no any prior knowledge was used, and the optimal angles were used to verify whether our nowGABAO could produce optimal beam angles or not when prior knowledge was incorporated.

### A. Clinical Head-and-Neck Case

A clinical head-and-neck (HN) case is shown in Fig. 3. A five-beam IMRT plan was chosen for this HN case, and the optimal beam angles previously found by exhaustive searching were 10°, 100°, 170°, 250° and 300° (Fig. 3). The dose prescription to the GTV was set to 76 Gy, which was normalized to 100%. Two angle constraints were defined for this case (Fig. 3): (a) 135°~155° and (b) 205°~225°, which reduced the discrete angle candidates in the search space from

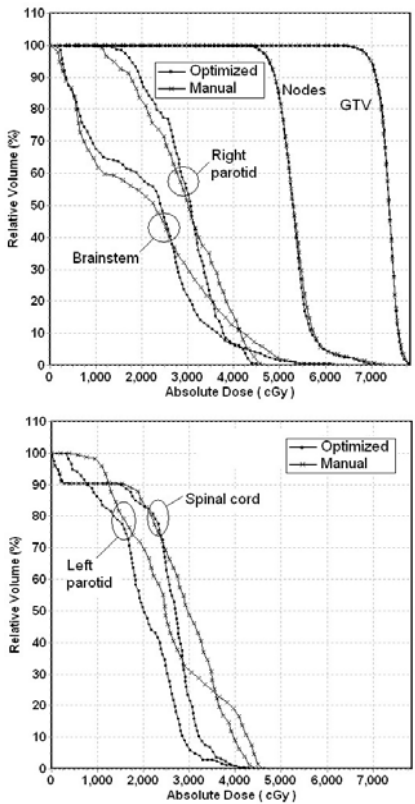

Fig. 5. Comparison between the DVHs produced by the manual plan (marked with crosses) and that produced by the nowGABAO optimized plan (marked with dots) for the head-and-neck case.

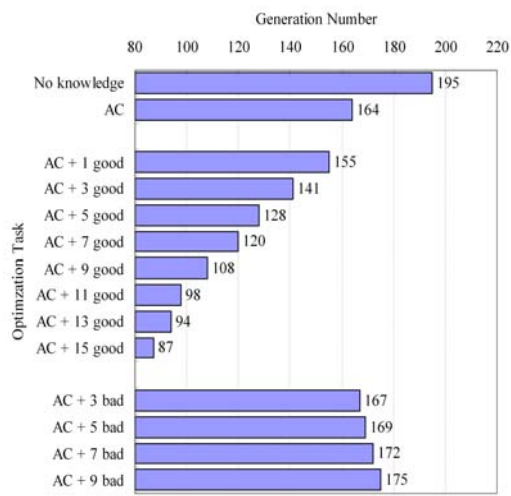

Fig. 6. Comparison of the generation numbers needed by different optimization tasks for the head-and-neck case. AC – angle constraint, good – good template, bad – bad template.

72 to 62. The good and bad plan templates were prepared in the way mentioned before.

Different nowGABAO runs with different knowledge combinations produced same optimized angles as the one given by exhaustive searching. Figure 4 compares the dose distributions of the nowGABAO optimized plan with the best manual plan among the 30 good plan templates. It is clear from the figure that the optimized plan provides better protection of OARs such as the parotid glands (numbered with '4' and '5') and spinal cord (numbered with '3') and a slightly better dose conformity to the GTV and nodes. The improvement in dose distributions can also be clearly observed in the DVHs compared in Fig. 5.

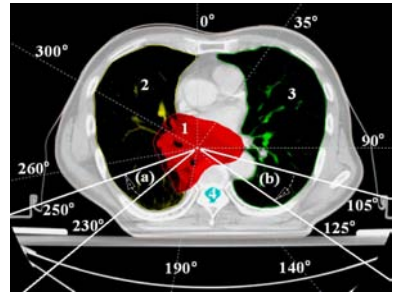

Fig. 7. The clinical lung case. There are four volumes involved: 1 – PTV, 2 – right lung, 3 – left lung, 4 – spinal cord. For a seven-coplanar-beam plan, the previously optimized beam angles were 0°, 35°, 90°, 140°, 190°, 260° and 300° (the light dotted straight lines). Two angle constraints were defined (the thick solid straight lines): (a) 105°~125° and (b) 230°~250°.

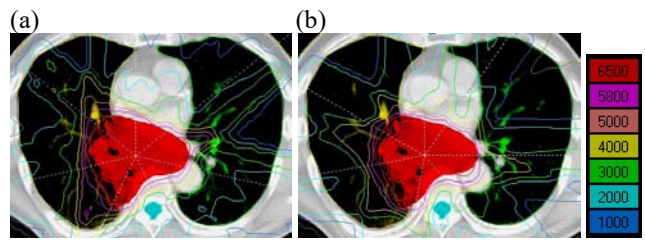

Fig. 8. Isodose contours of the lung case on a specific transverse for (a) the best manual plan among the plan template pool and (b) the optimized plan. The beam angles of the manual plan are 0°, 50°, 100°, 155°, 205°, 260° and 310° (the dotted straight lines in (a)). The optimized beam angles are 0°, 35°, 90°, 140°, 190°, 260° and 300° (the dotted straight lines in (b)). The legend of the isodose level (color vs. dose (cGy)) is shown at the right of (b).

The mean generation number over five runs for each optimization task is listed in Fig. 6, which clearly shows that good plan templates can dramatically reduce the needed generation numbers; whereas bad plan templates increase slightly the needed generation numbers, but this increasing is negligible if only a few bad templates were incorporated.

*B. Clinical Lung Case*

Figure 7 shows a clinical lung tumor case, which was designed to use a seven-beam IMRT plan. The optimal beam angles previously found by exhaustive searching were 0°, 35°, 90°, 140°, 190°, 260° and 300° (Fig. 7). The dose prescription to the PTV was set to 65 Gy, which was normalized to 100%. Two angle constraints were defined (Fig. 7): (a) 105°~125° and (b) 230°~250°. The good and bad plan templates were prepared in the way mentioned before.

Just as expected, all the nowGABAO tasks with different knowledge combinations produced same optimal beam angles.

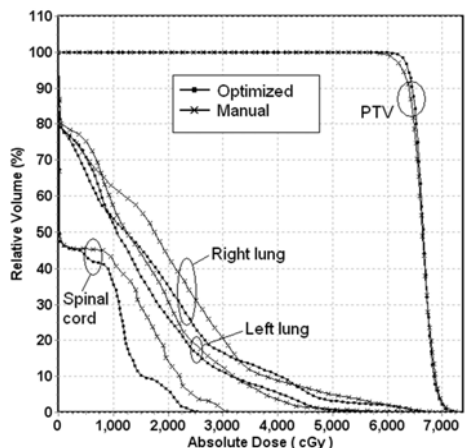

Fig. 9. Comparison between the DVHs produced by the best manual plan (marked with crosses) and that produced by the nowGABAO optimized plan (marked with dots) for the lung case.

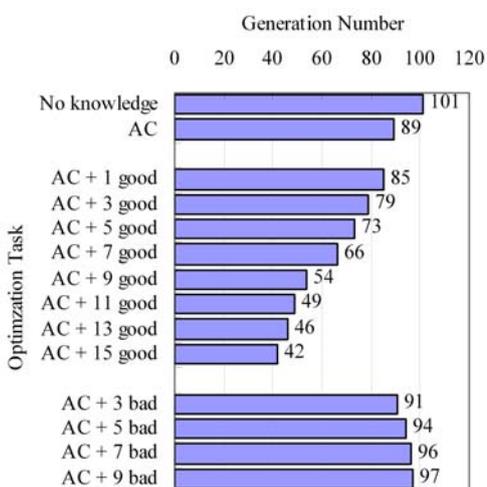

Fig. 10. Comparison of the generation numbers needed by different optimization tasks for the lung case. AC – angle constraint, good – good template, bad – bad template.

Both the isodose contours shown in Fig. 8 and the DVHs shown in Fig. 9 demonstrate the improvement in dose distribution produced by the nowGABAO optimized plan, which is reflected by improved dose conformity to the shape of the tumor (i.e. PTV) and reduced doses to the neighboring critical organs such as the spinal cord and left and right lungs.

Again, the generation numbers compared in Fig. 10 demonstrate a clear improvement in optimization efficiency contributed by good plan templates and a negligible degradation of optimization efficiency resulted from several bad templates.

### C. On the Computation Time

Both standard GA [20] and the proposed nowGABAO can find the optimal beam configuration, and the main purpose of nowGABAO is to improve the optimization efficiency. The computation time is obviously hardware dependant. The optimization tests in this study were undertaken with a personal computer (PC) with the configurations as: Intel(R) Core(TM)2 Duo CPU E7400 @ 2.80GHz, 2.0GB RAM. For the beam angle optimization using GA and CG, the computation costs are mainly on the beamlet intensity map optimization using CG method for each new individual (i.e. plan), whereas the computation costs on the genetic operations could be negligible. Also, the computation time used by each generation is case dependant. As for the HN and lung cases, the mean computation time per generation is about 21 s and 16 s, respectively. So for the two cases studied here, when the generation numbers were reduced from 195 and 101 to 87 and 42 after the defined angle constraints and 15 good templates were incorporated into the optimization, the computation time were reduced approximately from 69 min and 52 min to 31 min and 27 min. At the given situation of the cases and the parameter settings of the two studied cases, about 50% of the computation time was saved.

### IV. DISCUSSION AND CONCLUSION

Beam angle optimization (BAO) has become one of the subjects of intensive investigation in the field of radiation therapy, among which GA has been one of the popular algorithms to tackle this challenging problem [6], [12], [17], [20], [24], [30-32]. Based on the standard flowchart of GA for BAO [20], we proposed in this study a framework of prior knowledge and GA based beam angle optimization (nowGABAO) for IMRT planning. The prior knowledge of angle constraints contributes to the optimization efficiency by reducing the angle search space by excluding the defined angle range. As another type of knowledge, plan templates influence the optimization progress by initializing some of the individuals in the first generation of GA and replacing the worst individual in each new generation. The results on two clinical cases indicated that suitable defining of angle constraints and plan templates may obviously improve the optimization efficiency. Interestingly, we also found that a moderate number of quite bad plan templates have negligible influence on the degradation of optimization efficiency. Based on the extensive comparisons of various optimization tasks with different prior knowledge combinations, we believe that prior knowledge helps improve the optimization efficiency, and can be suitably incorporated by a GA due to its population based genetic operations and the elitist preservation based selection of GA.

The prior knowledge based framework for beam angle optimization proposed in this study is expected to be very desirable for clinical IMRT practice, because it is a relatively simple task for experienced planners to define suitable angle constraints and plan templates for most clinical cases, and they may have a practical choice to use their own experiences to help the computer optimizers. Furthermore, on the basis of the elitist preservation based selection operation in GA, nowGABAO with prior manual templates can guarantee that the quality of nowGABAO optimized plan is at least not worse than that of pre-defined manual plan templates when the optimizers fail to find the optimal angles due to some limitations of the optimizers, such as the unsuitable parameters specified for GA, though the failure has not occurred in this study.

As mentioned earlier, the BAO task in IMRT is more complicated and less intuitive due to the mutual compensations among the different beam angles as the result of computer optimized beamlet intensities [30]. This study was focused on the scheme of beam angle selection, and the beamlet intensity maps were optimized with the commonly used CG method [35]. Several investigators have studied in details the problem of being potentially trapped in local minima by using a local search method such as CG for beamlet intensity map optimization, and results show that those minima are very close to each other in cost function value and the resulting treatment plans are practically identical [36].

When the proposed idea is to be implemented in a commercial treatment planning system (TPS) in the future, the prior knowledge of plan templates could be defined by planners via a specially designed graphical user interface (GUI) embedded in the TPS. These prior templates could also be automatically filtered out from the beam configuration database of the TPS according to some specific filtering options [37]. In addition, some kinds of beam ranking indices, such as BEVD [14], [25] and BIPPS [34], could also be used to define suitable prior plan templates.